\documentclass[sigplan,screen,nonacm,10pt]{acmart}

\AtBeginDocument{%
  }

\setcopyright{acmlicensed}
\copyrightyear{2018}
\acmYear{2024}
\acmDOI{XXXXXXX.XXXXXXX}

\acmConference[SPLASH 2024]{Proceedings of the 2023 ACM SIGPLAN International Symposium on New Ideas, New Paradigms, and Reflections on Programming and Software}{October 20--25, 2024}{Pasadena, California, USA}
\acmISBN{978-1-4503-XXXX-X/18/06}

\graphicspath{ {./images/} }
\usepackage{float}
\usepackage{subcaption}
\usepackage{algorithm}
\usepackage{algpseudocode}
\usepackage{todonotes}
\usepackage{ulem}
\usepackage{listings}
\usepackage{xcolor} % For color support
\usepackage[english]{babel} % For better hyphenation
\usepackage{multirow}
\usepackage{tikz}
\usetikzlibrary{shapes.geometric}
\usetikzlibrary{shapes.symbols}
\usetikzlibrary{shapes.misc}

\theoremstyle{empty}

\newcommand\vtrue[1]{\overline{#1}}
\newcommand\vfalse[1]{\underline{#1}}
\newcommand\true{\mathsf{true}}
\newcommand\false{\mathsf{false}}

\newcommand\eval{\psi}
\newcommand\ring{\mathbb{Z}_{q}/(X^N+1)}
\newcommand\Ring{\mathcal{R}_q}
\newcommand\sext[2]{\left[{#1}\right]_{#2}}

\newcommand\enc{\mathsf{Enc}}
\newcommand\dec{\mathsf{Dec}}
\newcommand\encode{\mathsf{Encode}}
\newcommand\decode{\mathsf{Decode}}

\newcommand\tand{\mathsf{AND}}
\newcommand\txor{\mathsf{XOR}}
\newcommand\lwe[1]{\mathsf{LWE}(\Delta {#1})}
\newcommand\glwe[1]{\mathsf{GLWE}(\Delta {#1})}
\newcommand\sampext{\mathsf{Ext}}
\newcommand\keyswitch{\mathsf{KSw}}
\newcommand\blindrot{\mathsf{BRot}}

\newcommand\gateLUT[+1]{\mathsf{g_{#1}}}
\newcommand\gateLUTHomo{\mathsf{G}}
\newcommand\demap{\pi}
\newcommand\demapHomo{\Pi}
\newcommand\vect[1]{\mathbf{#1}}
\newcommand\dotprod[2]{\langle \vect{#1}, \vect{#2} \rangle}
\newcommand\nint[1]{\lfloor{#1}\rceil}
\newcommand\expected[1]{\mathbb{E}[#1]}
\newcommand\gauss[2]{\mathcal{N}({#1},{#2})}
\newcommand\erfc{\operatorname{erfc}}
\newcommand\mpv[1]{\mathsf{mpv}_{#1}}

\setcopyright{none}

\begin{document}

\title{ Free-XOR TFHE via Lazy Bootstrapping}


\author{Ali Ajorian}
\affiliation{%
 \institution{University of Basel}
 \country{Switzerland}}
\email{ali.ajorian@unibas.ch}

\author{Osman Biçer}
\affiliation{%
 \institution{University of Basel}
 \country{Switzerland}}
\email{osman.bicer@unibas.ch}

%\renewcommand{\shortauthors}{Ali Ajorian}

\begin{abstract}

Gate bootstrapping (GB) is a widely used technique in FHEW-like fully homomorphic encryption schemes, such as TFHE, for evaluating Boolean gates while simultaneously reducing the noise of the resulting ciphertext. However, the computational cost of GB is typically dominated by an expensive blind rotation (BR). \cite{chen2024free} proposed a gate bootstrapping framework in which $\txor$ gates can be evaluated simply by adding the two input ciphertexts, resulting in what they call \textit{free-XOR}. To obtain a functionally complete set of operations, they further designed a GB procedure for evaluating $\tand$ gates using two lightweight blind rotations. In this work, we propose a similar gate bootstrapping framework that supports free evaluation of $\txor$ gates while evaluating \textit{all symmetric binary Boolean gates}, i.e., the 8 gates whose output is identical for the inputs $(0,1)$ and $(1,0)$, using two blind rotations. In addition to providing broader gate support, our construction is substantially simpler than that of \cite{chen2024free}, both conceptually and in terms of implementation. As a proof of concept, we implemented our approach and compared it with standard TFHE, measuring speedups of about \textbf{30\%} for AES and \textbf{34\%} for SHA.
\end{abstract}




%\received{25 April 2024}

\maketitle

%\ajr{\today}

\section{Introduction}
Fully homomorphic encryption (FHE) allows computation directly on encrypted data without decrypting it. Among FHE schemes, TFHE~\cite{chillotti2020tfhe} is widely used for Boolean circuits. In the conventional TFHE approach, each bit is encrypted independently as an LWE ciphertext, and each gate is evaluated through \textit{programmable bootstrapping} (PBS), which computes a function on the encrypted value while refreshing the ciphertext to control noise accumulation. PBS is one of the most expensive operations in TFHE, so the cost of evaluating a circuit is strongly influenced by the number of bootstrappings it requires.

In conventional TFHE, every gate, including XOR gates, is evaluated via PBS. Chen et al.~\cite{chen2024free} proposed an approach called \textit{free-XOR}, which allows XOR gates to be evaluated without PBS. This is possible because XOR corresponds to addition modulo \(2\), which can be implemented directly using lightweight homomorphic addition. Therefore, circuits containing many XOR gates, such as those arising in symmetric cryptography and arithmetic, can benefit significantly from this technique.

Although achieving free-XOR for circuits consisting only of XOR gates is straightforward, extending the approach to support a complete set of Boolean gates is considerably more challenging. To obtain a functionally complete gate set, Chen et al.~\cite{chen2024free} proposed an algorithm for evaluating AND gates based on the NTRU hardness assumption, at the cost of two blind rotations.

\paragraph{Our contribution.}
In this work, we propose a similar gate bootstrapping framework that supports free evaluation of $\txor$ gates while evaluating \textit{all symmetric binary Boolean gates}, i.e., the 8 gates whose output is identical for the inputs $(0,1)$ and $(1,0)$, using two blind rotations. Our construction is conceptually simpler than that of \cite{chen2024free}, only based on learning with error (LWE) hardness assumption, making it easier to implement.

Our main ideas are as follows.
\begin{itemize}
\item \textbf{Two-bit encoding with the Free-XOR property.}
We encode each bit as an element of $\mathbb{Z}_4$, with the least significant bit representing the encoded Boolean value. Addition modulo $4$ therefore computes the XOR of the underlying bits without requiring a programmable bootstrapping (PBS). The most significant bit acts as a redundant component that can be eliminated through bootstrapping when necessary. As in conventional TFHE constructions, we also reserve a padding bit to support non-negacyclic functions during programmable bootstrapping. Additionally, we introduce headroom into the encoding to accommodate multiple consecutive XOR additions over the integers without carrying into the padding bit. Since the available headroom is bounded and the noise accumulates across successive additions, we periodically estimate headroom consumption and refresh the ciphertext when necessary. We refer to this strategy as \textit{lazy bootstrapping}.

\item \textbf{Symmetric gate evaluation with two PBS.}
For two input bits $b_0,b_1\in\{0,1\}$ of a symmetric gate, the corresponding truth-table row can be indexed by $i=2b_0+b_1\in\{0,1,2,3\}$. Given their two-bit encoded values $m_0$ and $m_1$, we obtain this index as
$i=2m_0+\pi(m_1)\pmod{4}$, where $\pi(x)=x\bmod 2$. The function $\pi$ is evaluated homomorphically using one PBS, yielding an encryption of the truth-table index. A second PBS then applies a lookup table representing the gate's truth table to obtain an encryption of the gate output. Thus, any two-input symmetric Boolean gate can be evaluated using two PBSs.

\item \textbf{Headroom analysis and failure probability.}
We analyze the constraints on the headroom, deriving a lower bound determined by the number of consecutive additions between bootstrappings and an upper bound imposed by modulus-switching noise. We also analyze the resulting failure probability to guide the selection of suitable parameters.
\end{itemize}

%\ajr{We implemented our approach on Rust using TFHE-rs low-level library \cite{TFHE-rs}. Our experiments show that our framework reduces the running time of circuits such as AES-128 by \textbf{25\%} compared to the standard TFHE gate bootstrapping.}
We implemented our approach in Rust using the low-level API of TFHE-rs \cite{TFHE-rs}. As expected, our experimental results show that our framework reduces circuit evaluation time compared with standard TFHE gate bootstrapping when the number of XOR gates significantly exceeds the number of symmetric gates. This improvement stems from evaluating XOR gates without programmable bootstrapping (PBS), whereas each symmetric gate requires two PBS operations. Consequently, our approach is particularly beneficial for circuits in which XOR gates dominate the gate composition. In particular, it speeds up the evaluation of AES by about \textbf{30\%} and SHA by about \textbf{34\%}.

\paragraph{Organization.}
Section~\ref{sec:background} reviews the necessary background, including LWE and GLWE encryption, programmable bootstrapping, and TFHE for Boolean circuits. Section~\ref{sec:freexor} presents our free-XOR framework, including the noise analysis. Section~\ref{sec:implementations} describes our implementation and experimental results. Finally, Section~\ref{sec:conc} summarizes our work.

%%%%%%%%%%%%%%%%%%%%%%%%%%%%%%%%%%%%%%%%%%%%% background %%%%%%%%%%%%%%%%%%%%%%%%%%%%%%%%%%%%%%%%%%%%%
\section{Background}
\label{sec:background}
In this section, we introduce the fundamental concepts used throughout the rest of the paper. In particular, we describe the Learning With Errors (LWE) problem and the LWE and GLWE encryption schemes employed by TFHE.

\subsubsection*{Notations}
When an integer $x$ encodes the Boolean value $\mathsf{false}$, we denote it by $\vfalse{x}$. Similarly, when $x$ encodes $\mathsf{true}$, we denote it by $\vtrue{x}$. We denote polynomials by uppercase letters, whereas \(X\) represents the indeterminate in polynomial expressions. Given a polynomial $P$, we denote by $\sext{P}{i}$ its coefficient of the monomial $X^i$. Vectors are denoted by bold characters, whereas scalars are written in non-bold characters. We use $a \gets A$ to denote sampling $a$ from a distribution or a set $A$. When $A$ is a set, the sampling is assumed to be uniform. 

We denote the polynomial ring $\ring$ by \(\Ring\). The expectation of a random variable \(X\) is denoted by \(\expected{X}\). A random variable \(X\) following a Gaussian distribution with mean \(\mu\) and variance \(\sigma^2\) is denoted by \(X\sim\gauss{\mu}{\sigma^2}\). For a standard Gaussian random variable \(Z\), we have 
$$\Pr(|Z|\ge l)=\erfc(l/\sqrt{2})$$
 where, $\erfc$ stands for complementary error function, defined as $\erfc(x)=1-\operatorname{erf}(x)$,  and $\operatorname{erf}$ is the error function defined by 
$$\operatorname{erf}(x)= \frac{2}{\sqrt{\pi}} \int_0^x e^{-t^2}\,dt.$$
We use $\nint{\cdot}$ to denote rounding to the nearest integer. The inner (dot) product of two vectors $\vect{a}$ and $\vect{b}$ is denoted by $\dotprod{a}{b}$.

\subsection{FHE}
Fully homomorphic encryption (FHE), often regarded as a long-standing goal of modern cryptography, is a framework in which a function $f$ can be evaluated directly on a ciphertext $c$ encrypting a message $x$, producing a ciphertext that decrypts to $f(x)$. The notion of computing on encrypted data was originally introduced as \textit{privacy homomorphism} by Rivest, Adleman, and Dertouzos in 1978~\cite{rivest1978data}, while the first construction supporting arbitrary-depth computation was given by Gentry nearly three decades later~\cite{gentry2009fullyphdthesis}. Before Gentry's construction, several homomorphic encryption schemes supported restricted classes of computations. For example, RSA~\cite{rivest1978method} and ElGamal~\cite{elgamal1985public} support a single homomorphic operation and are therefore referred to as \textit{partially homomorphic encryption} (PHE). Other schemes, such as BGN~\cite{boneh2005evaluating}, support both addition and multiplication, but only for circuits with a bounded multiplicative depth; these are commonly referred to as \textit{somewhat homomorphic encryption} (SHE). Surveys of homomorphic encryption and its different categories can be found in~\cite{acar2018survey,marcolla2022survey}.

A fundamental limitation of early SHE schemes is that homomorphic operations, particularly multiplication, introduce noise into ciphertexts. As this noise accumulates, it eventually exceeds the decryption tolerance, causing decryption to fail. Gentry addressed this limitation by introducing \textit{bootstrapping}, in which a noisy ciphertext is homomorphically decrypted using an encryption of the secret key, producing a new ciphertext encrypting the same plaintext with reduced noise. By periodically bootstrapping a ciphertext before its noise becomes too large, an encryption scheme can support arbitrarily deep computations and thereby achieve FHE.

An FHE scheme only needs to support a functionally complete set of operations to evaluate arbitrary functions. Based on the type of computation they natively support, FHE schemes can be broadly divided into three categories. First, \textit{Boolean FHE} schemes, such as TFHE~\cite{chillotti2016faster} and FHEW~\cite{ducas2015fhew}, operate on Boolean values and evaluate Boolean circuits composed of functionally complete sets of gates, such as NAND alone or the combination of AND and XOR. Second, \textit{arithmetic FHE} schemes, such as BGV~\cite{brakerski2014leveled} and BFV~\cite{brakerski2014efficient}, support modular addition and multiplication over finite rings and are particularly well suited to arithmetic circuits and SIMD-style batched computation. Third, \textit{approximate-arithmetic FHE} schemes, such as CKKS~\cite{cheon2017homomorphic}, support arithmetic over approximate representations of real or complex numbers and are designed for applications in which numerical approximation is acceptable, such as statistical and machine-learning computations.

FHE schemes can also be classified according to their key structure as either \textit{symmetric-key} or \textit{public-key} schemes. In particular, a symmetric-key FHE scheme can, under suitable constructions, be transformed into a public-key scheme~\cite{rothblum2011homomorphic}.

\subsection{TFHE}
Fast Fully Homomorphic Encryption over the Torus (TFHE) is a fully homomorphic encryption scheme known for its efficient and fast evaluation of Boolean circuits over encrypted data~\cite{chillotti2020tfhe}. Its security is based on the Learning With Errors (LWE) problem, introduced by Regev~\cite{regev2009lattices}. 

\subsubsection{LWE Hardness Asuumptions}
In its standard form, an LWE sample is generated as
$$b = \dotprod{a}{s} + e \pmod q,$$
where $\vect{a} \gets \mathbb{Z}_q^n$ is sampled uniformly at random, $\vect{s} \gets \mathbb{Z}_q^n$ is the secret vector\footnote{Usually, the secret vector is sampled as a binary vector $\vect{s}\gets\{0,1\}^n$.}, and $e \gets \chi$ is sampled from an error distribution $\chi$, typically a discrete Gaussian distribution $\gauss{0}{\sigma_e}$.

The hardness of LWE is commonly formulated in two variants:
\begin{itemize}
\item \textbf{Decision-LWE:} Given samples $(\vect{a},b)$ generated according to the LWE distribution and samples $(\vect{a},u)$, where $u \gets \mathbb{Z}_q$ is uniformly random, no PPT adversary can distinguish between the two distributions with more than a negligible advantage.
\item \textbf{Search-LWE:} Given polynomially many LWE samples $(\vect{a}_i,b_i)$ satisfying
$b_i = \dotprod{a}{s} + e_i \pmod q,$
recovering the secret vector $\vect{s}$ is assumed to be computationally hard.
\end{itemize}

\subsubsection{LWE Encryption Scheme}
\label{sec:lwe_enc}
To encrypt a message $m\in\mathbb{Z}_p$, where $p\ll q$, the plaintext must first be \textit{encoded} into $\mathbb{Z}_q$ using a scaling-based encoding:
$$\encode(m)=\Delta m+e, \qquad \Delta=\nint{\frac{q}{p}}$$
where typically $p|q$ and $e\gets\chi$ is an error. This encoding maps consecutive plaintext values in $\mathbb{Z}_p$ to points in $\mathbb{Z}_q$ separated by approximately $\Delta$. The resulting spacing provides room to place the error $e$ introduced by LWE\footnote{Each element of $\mathbb{Z}_q$ can be represented using $\log_2(q)$ bits, whereas a message in $\mathbb{Z}_p$ requires $\log_2(p)$ bits. The scaling effectively places the message in the most significant bits of the representation in $\mathbb{Z}_q$, leaving the remaining lower bits to accommodate the noise.}. 
This noisy encoding $\Delta m+e\pmod q$ is known as \textit{ciphertext phase}. Decoding this value gives
$$\decode(\Delta m+e)= \nint{\frac{\Delta m+e}{\Delta}} =m+\nint{\frac{e}{\Delta}}.$$

Since the error distribution $\chi$ has unbounded support, correct decoding is not guaranteed for every sample. Instead, $\sigma_e$ is chosen such that the probability of $|e|\geq \Delta/2$ is negligible. Consequently, $\nint{e/\Delta}=0$
with overwhelming probability, and the decoding correctly recovers $m$. Assuming  $e\sim\gauss{0}{\sigma_e^2}$ and letting $Z\sim\mathcal{N}(0,1)$,
\begin{align*}
P_{\mathrm{fail}}
&=\Pr\bigl(|e|\ge \Delta/2\bigr)
 =\Pr\!\left(|Z|\ge \frac{\Delta}{2\sigma_e}=l\right)
\end{align*}
and therefore $P_{\mathrm{fail}}=\erfc(l/\sqrt{2})$. This gives the decoding failure probability for the parameters $q$, $p$ and
$\sigma_e$.

The noisy-encoded plaintext is incorporated into an LWE ciphertext by adding the inner product of a uniformly random vector and a secret key:
$$\enc(\vect{s},m)=(\vect{a},b)=
\left(
\vect{a}\gets\mathbb{Z}_q^n \ , 
\dotprod{a}{s}+\Delta m+e
\pmod q
\right)
$$
where $\vect{s}$ is a secret key sampled uniformly at random from $\mathbb{Z}_q^n$. The vector $\vect{a}$ and the scalar $b$ are referred to as the \textit{mask} and \textit{body}, respectively. In the rest of the paper, we denote $\enc(\vect{s},m)$ by $\lwe{m}$ interchangeably.

Given a ciphertext $\lwe{m}=(\vect{a},b)$, decryption first removes the contribution of the secret key:
$$\dec(\vect{s},\vect{a},b)= b-\dotprod{a}{s}
=\Delta m+e \pmod q
$$
The plaintext is then recovered by applying the decoding function:
$$m=\decode(\Delta m+e)$$
which correctly recovers $m$ with overwhelming probability.

\subsubsection*{Homomorphic Operations} To add two messages homomorphically in the ciphertext space, it is sufficient to add their corresponding ciphertexts component-wise. Similarly, multiplying a plaintext scalar $\alpha$ by all components of a ciphertext corresponding to a message $m$ results in a ciphertext corresponding to $\alpha m$. Homomorphic multiplication of two ciphertexts, however, is not simply performed by component-wise multiplication of two LWE ciphertexts and requires additional techniques. We refer to~\cite{joye2021guide} for a detailed treatment of homomorphic operations.

A sequence of homomorphic operations causes the noise in a ciphertext to accumulate. As the noise increases, the resulting ciphertext may eventually violate the condition $|e|<\Delta/2$, making correct decryption impossible. To prevent this, a bootstrapping operation can be performed before the noise becomes too large. Bootstrapping reduces the noise in a ciphertext and produces a refreshed ciphertext that can be used for further homomorphic computation.

In TFHE, bootstrapping can additionally evaluate a function on the encoded plaintext while simultaneously reducing the ciphertext noise. This technique, known as \textit{programmable bootstrapping}, allows a single bootstrapping operation to perform both noise reduction and function evaluation. Programmable bootstrapping is discussed in Section~\ref{sec:pbs}.

\subsubsection{GLWE Encryption Scheme}
\label{sec:glwe}
The LWE encryption scheme can be generalized to the \textit{General Learning With Errors (GLWE)} scheme by replacing the underlying space $\mathbb{Z}_q$ with a polynomial ring $\Ring = \mathbb{Z}_q[X]/(X^N+1)$. In GLWE, the secret key $\vect{S}$ and the mask $\vect{A}$ are vectors of polynomials in $\mathcal{R}^n_q$, while the message $M$ and the error $E$ are polynomials in the same ring. Specifically, for a message $M\in\mathcal{R}_p$, where $p\ll q$, the encryption is defined as
$$
\enc(\vect{S},M)=(\vect{A},B)=
\left(
\vect{A}\gets\Ring^k\ ,
\dotprod{A}{S}+\Delta M+E
\right),
$$
where $\vect{S}\in\Ring^k$ is the secret key, $\Delta=\nint{q/p}$ is the scaling factor, and $E\in\Ring$ is an error polynomial whose coefficients are sampled from an error distribution $\chi$. In the rest of the paper, $\enc(\vect{S},M)$ is denoted by $\glwe{M}$ interchangeably.

Given a ciphertext $\glwe{M}=(\vect{A},B)$, decryption first removes the contribution of the secret key:
$$
\dec(\vect{S},\vect{A},B)=B-\dotprod{A}{S}
=\Delta M+E\pmod{q,\;X^N+1}.
$$
The resulting noisy encoded message is then decoded coefficient-wise by
$$\decode(\Delta M+E)=\nint{\frac{\Delta M+E}{\Delta}}.$$

In some applications, a polynomial $M$ needs to be represented in the form of a GLWE ciphertext without introducing encryption noise or hiding the polynomial. In this case, the mask is set to zero, i.e., $\vect{A}=\vect{0}$, and the body is set to $B=\Delta M$. This representation is called a \textit{trivial encryption} of $M$ and is denoted by$(\vect{0},\Delta M)$.

\subsubsection*{Sample Extraction}
Given a GLWE ciphertext $\vect{C}\in\Ring^{k+1}$ encrypting a polynomial
$$
M=\sum_{i=0}^{N-1}m_iX^i\in\ring,
$$
one can homomorphically construct an LWE ciphertext $\vect{c}\in\mathbb{Z}_q^{n+1}$, where $n=kN$, encrypting a selected coefficient $m_i$ of $M$. This can be achieved by appropriately selecting and rearranging coefficients of the GLWE ciphertext $\vect{C}$ to form the LWE ciphertext $\vect{c}$. Consequently, the secret key of $\vect{c}$, known as the \textit{extracted secret key}, is obtained by flattening the coefficients of the GLWE secret key:
$$\vect{s}=(s_{0,0},\ldots,s_{0,N-1},\ldots,
s_{k-1,0},\ldots,s_{k-1,N-1})
\in\mathbb{Z}_q^{kN}.
$$
\noindent This operation is called \textit{sample extraction} \cite{chillotti2020tfhe} and is denoted by
$$\sampext(\vect{C},i)= \lwe{m_i}\in\mathbb{Z}_q^{kN+1}.$$

\subsubsection*{Key Switching}
A key-switching operation is a homomorphic operation that takes as input a ciphertext $\mathsf{c}_{k_1}$, encrypted under a key $k_1$, together with an encryption of every component of $k_1$ under a (possibly different) key $k_2$, and produces a ciphertext $\mathsf{c}_{k_2}$ encrypting the same underlying message under $k_2$. The set of ciphertexts encrypting the elements of $k_1$ under $k_2$ is called the \textit{key-switching key} (KSK). The operation is denoted by:
$$\keyswitch(\mathsf{c}_{k_1}, \mathsf{KSK}) = \mathsf{c}_{k_2}.$$

The general idea is to homomorphically cancel $k_1$ out of the ciphertext using the encryption of $k_1$ under $k_2$, yielding a ciphertext under $k_2$. Several types of key switching are used in practice, including LWE-to-LWE, GLWE-to-GLWE, LWE-to-GLWE, and many-LWE-to-one-GLWE~\cite{chillotti2022tfhedeepdive}.

\subsubsection*{Blind Rotation}
\newcommand{\brpos}{t}

Homomorphic polynomial rotation has several applications, particularly in programmable bootstrapping. In the plaintext domain, rotating the coefficients of a polynomial $M$ by $\brpos$ positions corresponds to multiplying $M$ by $X^{-\brpos}$. When $M$ belongs to the polynomial ring
$\Ring=\ring$, this operation has two important properties:

\begin{enumerate}
    \item \textbf{Periodicity modulo $2N$.}
    Since $X^N\!=\!-1$, we have $X^{2N}\!=\!1$. Therefore, the rotation exponent is interpreted modulo $2N$, and we can restrict the rotation position to $0\leq\brpos<2N$.

    \item \textbf{Negacyclicity.}\footnote{A function $f$ is called negacyclic when for all $x\in D_f$ it satisfies $f(x+N) = -f(x).$}
    Since $X^N\!=\!-1$, $X^{-N}\!=\!-1$ as well\footnote{Since $X^{N}=-1$, we have $1=X^0 = X^{N-N} = X^N\cdot X^{-N}=-1\cdot X^{-N}=-X^{-N}$, and hence $X^{-N} = -1$. }. Consequently,
    \begin{align*}
        X^{-(t+N)}M(X)
        &=X^{-N}X^{-t}M(X)\\
        &=-X^{-t}M(X)
        \pmod{X^N+1}.
    \end{align*}
    Thus, rotating by an additional $N$ positions negates the resulting polynomial.
\end{enumerate}

Given a GLWE ciphertext $\vect{C}$ encrypting a polynomial $M\in\Ring$ and an LWE ciphertext $\vect{c}=(\vect{a},b)$ encrypting a rotation position $t\in\{0,\ldots,2N-1\}$ under the secret key $\vect{s}$, TFHE homomorphically computes a GLWE ciphertext $\vect{C}'\in\Ring^{k+1}$ encrypting the rotated polynomial. This operation is called \textit{blind rotation} because the rotation is performed without revealing the encrypted position $t$. It relies on GLWE encryptions of the individual secret-key components $s_i$, which collectively form the \textit{blind rotation key} (BRK). The operation is denoted by
\[
    \blindrot(\mathsf{BRK},\vect{C},\vect{c})
    =\vect{C}'\in\Ring^{k+1}.
\]

To understand how blind rotation is performed, recall that the LWE ciphertext satisfies
\[
    b-\sum_{i=0}^{n-1}a_i s_i
    =\Delta t+e\pmod q,
\]
The rotation position is therefore represented by a noisy encoding in $\mathbb{Z}_q$, rather than being available directly as a plaintext integer. TFHE uses modulus switching to map the ciphertext components from $\mathbb{Z}_q$ to the rotation-exponent domain modulo $2N$. At the algebraic level, the underlying decomposition is
\[
    X^{\sum_{i=0}^{n-1}a_i s_i-b}M
    =X^{-b}M\prod_{i=0}^{n-1}X^{a_i s_i}.
\]
This decomposition consists of an initial rotation determined by $-b$, followed by conditional rotations: for each $i$, the polynomial is rotated by $a_i$ when $s_i=1$ and remains unchanged when $s_i=0$. The BRK enables these conditional rotations to be evaluated homomorphically without revealing the secret-key bits. Blind rotation is therefore a central and computationally expensive operation in TFHE.

\subsection{Programmable Bootstrapping (PBS)}
\label{sec:pbs}
Introduced by Gentry~\cite{gentry2009fullyphdthesis}, bootstrapping reduces the noise accumulated in a ciphertext through a sequence of homomorphic operations by homomorphically evaluating its decryption function, using a homomorphic encryption of the secret key. The resulting ciphertext encrypts the same underlying plaintext, while its noise is essentially independent of the noise in the input ciphertext. TFHE further allows a target function $h$ to be evaluated during bootstrapping, i.e.,
$$\mathsf{PBS}(\lwe{m})=\lwe{h(m)},$$
by encoding $h$ into a polynomial called a \textit{test polynomial}. We first elaborate on test polynomials in Section~\ref{sec:lut} and then provide further details on the operation of PBS in Section~\ref{sec:pbs_details}.

\subsubsection{Test Polynomial and Lookup Table (LUT)}
\label{sec:lut}
Consider a polynomial $V\in \ring$ as
$$V(x) = \sum_{i=0}^{N-1}v_iX^i, \quad v_i\in \mathbb{Z}_q$$
and define the following function for each $t\in \mathbb{Z}_{2N}$
\begin{align*}
\eval_{V}(t)&\triangleq  \sext{X^{-t}V(X)}{0} \pmod{X^N+1}
\end{align*}
That is, $\eval_V(t)$ rotates $V$ by $t$ positions via multiplying it by the monomial $X^{-t}$ and extracts the constant coefficient of the result. In the ciphertext domain, $V$ is represented as a GLWE trivial encryption, the multiplication is performed via a blind rotation, and the constant coefficient is extracted via sample extraction.

As discussed in the context of blind rotation, the operation $X^{-t}V(X)$ in $\Ring$ is negacyclic, therefore $\eval_{V}(t)$ for $t<N$ samples the $t$-th coefficient of $V$, while for $t\geq N$ it samples the negative of the coefficient at position $t-N$, namely:
$$ \eval_V(t) =\begin{cases} v_t \quad\quad\quad 0\leq t <N \\ -v_{t-N} \quad\ N\leq t <2N\end{cases}$$
Note that the coefficients $v_0, \ldots, v_{N-1}$ of $V$ remain free and unconstrained, while the generated length-$2N$ sequence $\eval_V(0), \eval_V(1),\ldots, \eval_V(2N-1)$ is negacyclic, with its second half equal to the negation of its first half.

A negacyclic function $f:\mathbb{Z}_{2N}\to \mathbb{Z}_q$ can be encoded by $\eval_V$ using the polynomial
\begin{equation}
\label{eq:test_poly}
V(X) = \sum_{i=0}^{N-1} f(i)X^i
\end{equation}
which satisfies $f(m)=\eval_V(m)$ and can be computed homomorphically via a blind rotation followed by a sample extraction. Such a polynomial is called a \textit{test polynomial} and can be represented as a lookup table containing $N$ entries, each corresponding to one of its coefficients. We refer to such a function $f$ as \textit{LUT function}.

Negacyclicity is a structural property of $f$ and cannot be circumvented: a non-negacyclic function on domain $\mathbb{Z}_{2N}$ has no test-polynomial encoding. However, any function $g:D\to \mathbb{Z}_q$ with $D\subseteq \mathbb{Z}_N$ can still be encoded, by defining a negacyclic $f:\mathbb{Z}_{2N}\to \mathbb{Z}_q$ as
\begin{equation}
\label{eq:nonNegcyc}
f(i) \triangleq\begin{cases} g(i) & i \in D \\ -g(i\!-\!N) & i-N \in D\\ 0 & \text{o.w.}\end{cases}
\end{equation}
This works precisely because the target domain $D\subseteq \mathbb{Z}_N$ never reaches the second half of $\mathbb{Z}_{2N}$. Consequently, under this construction every valid input $m$ to $f$ has its most significant bit equal to zero — so at the design level, one padding bit must be reserved for the input $m$, and no sequence of operations may ever carry into it.

\subsubsection*{Recoding Input via Modulo Switching}
To evaluate $\eval_V(t)$ on an LWE ciphertext encoding $\Delta m+e\in\mathbb{Z}_q$, its value must be recoded into the index range $[0,2N)$. Direct reduction modulo $2N$ is unsuitable because it retains low-order bits, which primarily contain noise, while discarding the high-order bits encoding the message. Instead, we use \textit{modulus switching}~\cite{brakerski2014efficient,chillotti2022tfhedeepdive} to change the modulus from $q$ to $2N$.

Given an LWE ciphertext $(a_1,\ldots,a_n,b)$, modulus switching scales and rounds each component:
$$(\tilde a_1, \ldots, \tilde a_n, \tilde b) = \left(\nint{a_1\frac{2N}q} , \ldots ,\nint{a_n \frac{2N}q}, \nint{b\frac{2N}q}\right)$$
Consequently, the encoded message is rescaled from $\Delta=q/p$ to $\tilde{\Delta}=2N/p$:
$$(\Delta m + e)\frac{2N}{q}=
\left(\frac{q}{p}m+e\right)\frac{2N}{q}
=\frac{2N}{p}m + \frac{2N}{q}e=\tilde\Delta m + e',
$$
where $e'=(2N/q)e$ is the scaled noise, ignoring rounding errors. Thus, modulus switching preserves the message encoding while mapping it to the desired index range.

%\subsubsection*{Noise introduced by modulus switching.}
 Denoting rounding errors by $\varepsilon_i=\tilde a_i-\frac{2N}{q}a_i$ and
$\varepsilon_b=\tilde b-\frac{2N}{q}b$, the phase of the
switched ciphertext is
$$
\tilde b-\sum_{i=1}^n s_i\tilde a_i
=\tilde\Delta m+e'+e_{\mathrm{ms}}, \qquad
e_{\mathrm{ms}}=\varepsilon_b-\sum_{i=1}^n s_i\varepsilon_i .
$$

Each rounding error is modeled as independent and uniform on $[-1/2,1/2]$, so it has mean $0$ and variance $1/12$. Hence,
\begin{align*}
\mathrm{Var}(e_{\mathrm{ms}}) 
= \mathrm{Var}\left(\varepsilon_b-\sum_{i=1}^n s_i\varepsilon_i\right)
&=  \mathrm{Var}(\varepsilon_b) +  \sum_{i:\, s_i=1} \mathrm{Var}(\varepsilon_i)\\
&= \frac{1+u}{12}
\end{align*}
where $u$ denotes the number of elements in the secret key for which $s_i=1$. For a uniformly random binary key, $\expected{u}=n/2$, hence
$$
\sigma_{\mathrm{ms}}^2\approx\frac{n+2}{24}.
$$
The error $e'$ already present before the rounding has standard deviation $\frac{2N}{q}\sigma_e$,
where $\sigma_e$ is the standard deviation of $e$. Since $e'$ and $e_{\mathrm{ms}}$
are independent, the total error $E=e'+e_{\mathrm{ms}}$ seen by $\eval_V$ has standard deviation
\begin{equation}
\label{eq:sigma_total}
\sigma_E=\sqrt{\Bigl(\tfrac{2N}{q}\sigma_e\Bigr)^2+\frac{n+2}{24}}.
\end{equation}

Consecutive messages are $\tilde\Delta=2N/p$ positions apart, so decoding is correct
if and only if $|E|<\tilde\Delta/2=N/p$. Modeling the total error as Gaussian distribution $E\sim\gauss{0}{\sigma_E^2}$\footnote{By the central limit theorem, since \(E\) is a sum of several independent random variables, namely \(E=e'+\varepsilon_b-\sum_{i=1}^n s_i\varepsilon_i\), it can be approximated by a Gaussian distribution.} and denoting $Z=\frac{E}{\sigma_E}\sim\gauss{0}{1}$, the failure probability is
$$P_{\mathrm{fail}} =\Pr\left(|E|\ge\frac{N}{p} \right) =\Pr\left(|Z|\ge l\right), \qquad l=\frac{N}{p\sigma_E}$$
Since $P_{\mathrm{fail}}=\operatorname{erfc}(l/\sqrt{2})$, a target negligible failure probability determines $l$. For given $N$, $n$, and $\sigma_e$, the plaintext modulus must therefore satisfy
\begin{equation}
\label{eq:choosing_p}
\frac{N}{p\sigma_E} \geq l
\quad\Longrightarrow\quad
p\leq\frac{N}{l\sigma_E}.
\end{equation}

\subsubsection*{Designing Test Polynomials}
At the design level, we aim to evaluate $h:\mathbb{Z}_p\to\mathbb{Z}_p$, whereas $\eval_V$ implements $f:\mathbb{Z}_{2N}\to\mathbb{Z}_q$. This mismatch is addressed by adjusting the input and output encodings.

Since $p\ll N$, each plaintext $m\in\mathbb{Z}_p$ corresponds to a block of $\tilde{\Delta}=2N/p$ consecutive indices in $\mathbb{Z}_{2N}$. The input message associated with an index $t$ is therefore
\[
m=\nint{\frac{t}{\tilde{\Delta}}}
=\nint{\frac{t}{2N/p}}.
\]
To ensure valid ciphertext encoding, each output $h(m)$ is scaled by $\Delta$.

Accordingly, the $2N$ indices are partitioned into $2p$ blocks of $\tilde{\Delta}=N/p$ elements each, with every index in a block assigned the same scaled output. For a function compatible with negacyclic evaluation, the corresponding test polynomial is
\begin{equation}
\label{eq:nega_h}
V(X)=\sum_{i=0}^{N-1}
\Delta h\left(\nint{\frac{i}{N/p}}\right)X^i.
\end{equation}
Non-negacyclic functions can be handled using the technique described in Equation~\ref{eq:nonNegcyc}.

\subsubsection{How PBS Works}
\label{sec:pbs_details}
To bootstrap a ciphertext $\vect{c}=\lwe{m}$ encrypted under a secret key $\vect{s}=(s_1,\ldots,s_n)$ and containing accumulated error $e$, TFHE requires a \textit{bootstrapping key} (BSK). The BSK is actually a blind rotation key consists of a collection of GLWE ciphertexts under an encryption key $\vect{S}'$, with one ciphertext associated with each component of the input LWE secret key. These ciphertexts enable the components $s_i$ of the input secret key to be used homomorphically during the blind rotation. We refer to \cite{chillotti2022tfhedeepdive} for details on the construction of the BSK.

Given $\vect{c}$, the BSK, and a trivial GLWE encryption of a test polynomial $V$ corresponding to a target negacyclic function $h$\footnote{One can instead hide $V$ by encrypting it as a non-trivial GLWE ciphertext.}, TFHE homomorphically evaluates $\eval_V(t)$, where $t=(2N/q)\cdot(\Delta m+e)$,
to obtain a ciphertext $\lwe{h(m)}$. This proceeds as follows:
\begin{enumerate}
\item Obtain $\lwe{t}$ via modulus switching, i.e., by scaling all components of $\vect{c}$ by $2N/q$ and rounding the results.

\item Compute $X^{-t}V(X)$ homomorphically via a \textit{blind rotation}, which takes $\glwe{V}$, $\lwe{t}$, and the BSK as input and outputs a GLWE encryption of $X^{-t}V(X)$, denoted by $\vect{C'}$. In otherwords,
$$\vect{C'}=\blindrot(\mathsf{BSK},\vect{C},\vect{c})$$
This ciphertext is encrypted under the key $\vect{S}'$ by which BSK is generated.

\item Extract the constant coefficient of $\vect{C}'$ as an LWE ciphertext, yielding an LWE encryption of $h(m)$, via sample extraction operator $\sampext(\vect{C'},0)=\vect{c'} \in \mathbb{Z}_q^{kN+1}$. As discussed in Section~\ref{sec:glwe}, this extracted ciphertext is under a key $\vect{s'}\in \mathbb{Z}_q^{kN}$ obtained by flattening of  $\vect{S'}$.

\item Having an appropriate LWE to LWE key-switching key $KSK$ from $\vect{s'}$ to $\vect{s}$, switch the encryption key of $\vect{c'}$ from $\vect{s'}$ back to $\vect{s}$ via a  \textit{key-switching} operation as
$$\keyswitch(\vect{c'}, KSK) = \vect{c''} \in \mathbb{Z}_q^{n+1}$$
$\vect{c''}$ will be $\lwe{h(m)}$.
\end{enumerate}

Note that no modulus switching from $2N$ back to $q$ is needed. Although the input to the blind rotation is represented modulo $2N$, the test polynomial $\glwe{V}$ itself lives in $\mathbb{Z}_q[X]/(X^N+1)$ throughout the computation. Consequently, the sample-extracted LWE ciphertext is already defined modulo $q$ and is therefore a valid LWE ciphertext.

\subsubsection{TFHE Boolean Circuits}
In TFHE, Boolean values are encrypted as LWE ciphertexts, and binary gates can be evaluated homomorphically by applying a programmable bootstrapping (PBS) to an appropriate linear combination of the input ciphertexts. The linear combination depends on the Boolean function to be evaluated, while the PBS evaluates a corresponding look-up table (LUT) and simultaneously reduces the ciphertext noise. This process is commonly referred to as \textit{gate bootstrapping}. In particular, all standard binary Boolean gates, such as AND, OR, XOR, NAND, and NOR, can be evaluated in this way \cite{chillotti2020tfhe}.

A natural starting point for Boolean values is to set $p=2$ for a chosen ciphertext modulus $q$. According to the encoding introduced in Section~\ref{sec:lwe_enc}, this gives $\Delta=q/p=q/2$ and the two Boolean messages are encoded as $\encode(m)=\Delta m$, that is, $\encode(0)=0$ and $\encode(1)=q/2$. However, the linear combination of two such encoded values needs at least two bits of precision to remain distinguishable, so a larger $p$ must be chosen. Moreover, the underlying function $h:\mathbb{Z}_p \to \mathbb{Z}_p$ whose test polynomial realizes a given gate type is not negacyclic, so, as stated in Section~\ref{sec:pbs}, padding bits are needed.

An alternative, the \textit{shifted encoding} \cite{chillotti2016faster, chillotti2020tfhe, chillotti2022tfhedeepdive}, is defined as
$$\encode(m)=-\frac{q}{8} + \frac{\Delta}2 m, \quad \Delta = q/2,$$
which gives $\encode(0)=-q/8$ and $\encode(1)=q/8$. Under this encoding, boolean values take the form of a negative/positive sign representation, which makes gate bootstrapping more convenient. Here the LUT function is a sign function on the rotation index,
$$\mathsf{sgn}(x)=\begin{cases}-\dfrac{q}{8} & x<0 \\[4pt] +\dfrac{q}{8} & x\geq 0,\end{cases}$$
whose corresponding test polynomial is $V(X) = \sum_{i=0}^{N-1}\frac{q}{8}X^i$. Since $\mathsf{sgn}$ is negacyclic, no padding bit is required. It is then sufficient to define, for each gate type, a linear combination of the encoded inputs that $\mathsf{sgn}$ maps to the correct output. For example, as shown in Table~\ref{tbl:zama_or}, the combination $\encode(u)+\encode(v) + q/8$ evaluates the OR gate: $\mathsf{sgn}(-q/8)=-q/8$ encodes $0$, while $\mathsf{sgn}(q/8)=\mathsf{sgn}(3q/8)=q/8$ encodes $1$.

\begin{table}[H]
\centering
\begin{tabular}{cc|cc|ccc}
$u$ & $v$ & $U\!=\! \encode(u)$& $V\!=\!\encode(v)$&  $U\!+\!V\! +\! q/8$\\ \hline
$\false$ & $\false$ & $-q/8$ & $-q/8$ & $-q/8$ \\
$\false$ & $\true$ & $-q/8$ & $+q/8$ & $+q/8$\\
$\true$ & $\false$ & $+q/8$ & $-q/8$ & $+q/8$\\
$\true$ & $\true$ & $+q/8$ & $+q/8$ & $+3q/8$\\
\end{tabular}
\caption{Linear combination used for OR gate bootstrapping with the shifted Boolean encoding.}
\label{tbl:zama_or}
\end{table}

Similarly, the combination $\encode(u)+\encode(v) - q/8$ is the right choice for AND gates, as shown in Table~\ref{tbl:zama_and}.

\begin{table}[H]
\centering
\begin{tabular}{cc|cc|ccc}
$u$ & $v$ & $U\!=\! \encode(u)$& $V\!=\!\encode(v)$&  $U\!+\!V\! -\! q/8$\\ \hline
$\false$ & $\false$ & $-q/8$ & $-q/8$ & $-3q/8$ \\
$\false$ & $\true$ & $-q/8$ & $+q/8$ & $-q/8$\\
$\true$ & $\false$ & $+q/8$ & $-q/8$ & $-q/8$\\
$\true$ & $\true$ & $+q/8$ & $+q/8$ & $+q/8$\\
\end{tabular}
\caption{Linear combination used for AND gate bootstrapping with the shifted Boolean encoding.}
\label{tbl:zama_and}
\end{table}

%%%%%%%%%%%%%%%%%%%%%%%%%%%%%%%%%%%%%%%%%%%%% Free XOR %%%%%%%%%%%%%%%%%%%%%%%%%%%%%%%%%%%%%%%%%%%%%

\section{Free-$\txor$ Gate Bootstrapping Framework}
\label{sec:freexor}
For a circuit consisting only of XOR gates, an encoding with $p=2$ is sufficient, as XOR can be obtained simply by homomorphically adding the two input ciphertexts. This works because XOR is equivalent to addition modulo $2$, which the encoding preserves automatically, without requiring bootstrapping to evaluate nonlinear operations. Moreover, the noise grows relatively slowly, so many common circuits can be evaluated without bootstrapping for noise reduction. Specifically, after $k$ XORs the ciphertext contains $k+1$ independent encryption noises, each with standard deviation $\sigma_e$, so the total noise $E$ is approximately Gaussian with $\sigma_E=\sqrt{k+1}\,\sigma_e$. Decryption fails when $E$ is at
least half a step, $\Delta/2=q/(2p)=q/4$, hence
\begin{align*}
P_{\mathrm{fail}}
&=\Pr(|E|\ge \Delta/2) =\Pr\!\left(|Z|\ge q/4\sigma_E\right)\\
 &=\Pr\!\left(|Z|\ge l \right), \qquad l=\frac{q}{4\sqrt{k+1}\,\sigma_e}
\end{align*}
where $Z\sim\gauss{0}{1}$. For a target failure probability $2^{-\alpha}$, let $z_\alpha$ be the value satisfying $\Pr(|Z|\ge z_\alpha)=2^{-\alpha}$.
Requiring $l\ge z_\alpha$ gives the maximal number of XORs,
\begin{equation}
l\ge z_\alpha \implies 
k+1\le\left(\frac{q}{4\,z_\alpha\,\sigma_e}\right)^2
\end{equation}

\paragraph{Example.}
Take the modulus $q=2^{64}$ and a noise standard deviation $\sigma_e=2^{47}$, which is of the order commonly used for LWE ciphertexts with a dimension of a few hundred\footnote{It is common to give $\sigma_e$ as a fraction of the modulus $q$; here $\sigma_e/q=2^{-17}$.}. Then $q/(4\sigma_e)=2^{64}/(2^{2}\cdot 2^{47})=2^{15}$ and the bound becomes
$$k+1\le\left(\frac{2^{15}}{z_\alpha}\right)^2
\approx\frac{1.07\times10^{9}}{z_\alpha^{2}}.$$
For $\alpha=40$ we have $z_\alpha\approx 7.2$, which gives
$$k+1\le\frac{1.07\times 10^{9}}{7.2^{2}}\approx 2\times10^{7}.$$
Thus about twenty million XORs can be chained while keeping the failure probability below $2^{-40}$. For $\alpha=64$ ($z_\alpha\approx9.1$) the bound is about $1.3\times10^{7}$.

\subsection{Supporting Other Boolean Gate Types}
For circuits with other types of binary gates the situation is different. A linear combination of two input ciphertexts, which TFHE handles very efficiently, cannot implement a non-linear gate such as AND or OR.
Instead, the gate is implemented by a programmable bootstrapping applied to an appropriate linear combination of the inputs.

For bits \(b_0,b_1\in\{0,1\}\), the linear combination
\begin{equation}
\label{eq:id_plain}
i(b_0,b_1) = 2b_1+b_0\in\{0,1,2,3\}
\end{equation}
uniquely identifies the input pair, as shown in Table~\ref{tbl:id_unique}. Thus, the index $i$ identifies the corresponding row in the truth table of the target gate. This design supports only eight of the sixteen possible binary Boolean gates: AND, NAND, OR, NOR, XOR, XNOR, FALSE, and TRUE. These gates produce the same output for the input pairs $(0,1)$ and $(1,0)$. Consequently, their outputs can be determined from either index convention, $i=2b_1+b_0$ or $i=2b_0+b_1$. The remaining eight gates distinguish these two input pairs and are therefore not supported by our design.

 For a gate type $\textsf{T}$, we define a function $\gateLUT{T}:\mathbb{Z}_4\to\mathbb{Z}_2$ whose output matches the truth table of that gate, i.e.,  $T(b_0,b_1) = \gateLUT{T}(i(b_0,b_1))$. Homomorphically evaluating $i$ and then applying programmable bootstrapping with a lookup table (LUT) representing $\gateLUT{T}$ yields the correct gate output. We call $\gateLUT{T}$ the \textit{gate characteristic function} for a gate $\textsf{T}$. 

\begin{table}[H]
\centering
\begin{tabular}{cc|c|cc}
$b_1$&$b_0$&$i=2b_1+b_0$&$\gateLUT{AND}$ & $\gateLUT{OR}$\\ \hline
$0$ & $0$ &$0$ & $0$ & $0$\\
$0$ & $1$ &$1$& $0$ & $1$\\
$1$ & $0$ &$2$& $0$ & $1$\\
$1$ & $1$ &$3$& $1$ & $1$\\
\end{tabular}
\caption{Truth-table indices and corresponding output functions for AND and OR gates.}
\label{tbl:id_unique}
\end{table}

%%%%%%%%%%

\subsubsection{Encoding}
\label{sec:our_encoding}
We need a mapping\footnote{We avoid the term \textit{encoding} here because TFHE uses it to refer to the scaling-by-\(\Delta\) operation.} for each bit $b$ that supports Free-XOR while also allowing the index $i$ in Equation~\ref{eq:id_plain} to be computed efficiently. A two-bit mapping whose least significant bit equals $b$ achieves the Free-XOR property and, with programmable bootstrapping, also supports the computation of $i$. Specifically, $b=0$ is mapped to the bit patterns $00$ and $10$, denoted by $\vfalse{0}$ and $\vfalse{2}$, respectively, whereas $b=1$ is mapped to $01$ and $11$, denoted by $\vtrue{1}$ and $\vtrue{3}$, respectively. Formally, the mapping is given by $\demap:\mathbb{Z}_4\to\mathbb{Z}_2$, where
$\demap(m)=m \bmod 2$ is the parity of $m$, i.e., whether $m$ is even or odd. With a slight abuse of notation, we denote the canonical representatives $\vfalse{0}$ and $\vtrue{1}$ by $\demap^{-1}(0)$ and $\demap^{-1}(1)$ respectively.

We define a function \(I:\mathbb{Z}_4^2\to\mathbb{Z}_4\) that computes the index \(i\) from Equation~\ref{eq:id_plain} for any pair of mapped inputs:
\begin{equation}
\label{eq:func_I}
I(m_0,m_1)\triangleq 2m_0+\demap(m_1)\pmod 4,
\end{equation}
where \(\demap(x)=x\bmod 2\). Thus, for a gate type \(\mathsf{T}\) with characteristic function \(\gateLUT{T}\), the output for encoded inputs \(m_0,m_1\in\mathbb{Z}_4\) is given by
\begin{equation}
\label{eq:evaluation}
\mathsf{T}(\demap(m_0),\demap(m_1))=\gateLUT{T}(I(m_0,m_1)).
\end{equation}

To demonstrate the Free-XOR property and compatibility with Equation~\ref{eq:id_plain}, we introduce a table over the mapping space that we call the \textit{modulo table}. It contains 16 rows, one for each pair of values \(m_0,m_1\in\mathbb{Z}_4\). Each input pair in the Boolean truth table corresponds to four rows in the modulo table, since each bit has two possible representations. We call these four rows an \textit{equivalent chunk}. For example, the input pair \((1,0)\) corresponds to the four pairs
\[
(\vtrue{1},\vfalse{0}),\quad
(\vtrue{1},\vfalse{2}),\quad
(\vtrue{3},\vfalse{0}),\quad
(\vtrue{3},\vfalse{2}).
\]
Accordingly, the modulo table is partitioned into four equivalent chunks, denoted by \(Q_0,Q_1,Q_2,Q_3\), corresponding to the Boolean input pairs \((0,0),(0,1),(1,0),(1,1)\), respectively.

Table~\ref{tbl:fxor} shows that addition modulo \(4\) preserves the Free-XOR property for all 16 pairs of mapped values \(m_0,m_1\in\mathbb{Z}_4\). It also shows that the function \(I\) computes the index \(i\) from Equation~\ref{eq:id_plain}.
\begin{table}[t]
\centering
\setlength{\tabcolsep}{3pt}
\caption{The mapping $\demap$ achieves free-XOR by addition modulo $4$. The index $i$ is obtained as $I(m_0,m_1)=2m_0+\demap(m_1)$, where $\demap(m)=m\bmod 2$.}
\label{tbl:fxor}
\begin{tabular}{ccc|ccc|c}
$ m_0$& $ m_1$  & $m_0\!+\!m_1$ & $\demap(m_0)$ & $\demap(m_1)$ & $\demap(m_0\!+\!m_1)$ & $2m_0\!+\!\demap(m_1)$ \\ \hline
$\vfalse{0}$ & $\vfalse{0}$ &$\vfalse{0} $ &$0$& $0$& $0$ &$0$\\
$\vfalse{0}$ & $\vfalse{2}$ &$\vfalse{2} $ &$0$& $0$& $0$ &$0$\\
$\vfalse{2}$ & $\vfalse{0}$ &$\vfalse{0} $ &$0$& $0$& $0$ &$0$\\
$\vfalse{2}$ & $\vfalse{2}$ &$\vfalse{0} $ &$0$& $0$& $0$ &$0$\\
\hline
$\vfalse{0}$ & $\vtrue{1}$ &$\vtrue{1} $ &$0$& $1$& $1$ &$1$\\
$\vfalse{0}$ & $\vtrue{3}$ &$\vtrue{3} $ &$0$& $1$& $1$ &$1$\\
$\vfalse{2}$ & $\vtrue{1}$ &$\vtrue{3} $ &$0$& $1$& $1$ &$1$\\
$\vfalse{2}$ & $\vtrue{3}$ &$\vtrue{1} $ &$0$& $1$& $1$ &$1$\\
\hline
$\vtrue{1}$ & $\vfalse{0}$ &$\vtrue{1} $ &$1$& $0$& $1$ &$2$\\
$\vtrue{1}$ & $\vfalse{2}$ &$\vtrue{3} $ &$1$& $0$& $1$ &$2$\\
$\vtrue{3}$ & $\vfalse{0}$ &$\vtrue{3} $ &$1$& $0$& $1$ &$2$\\
$\vtrue{3}$ & $\vfalse{2}$ &$\vtrue{1} $ &$1$& $0$& $1$ &$2$\\
\hline
$\vtrue{1}$ & $\vtrue{1}$ &$\vfalse{2} $ &$1$& $1$& $0$ &$3$\\
$\vtrue{1}$ & $\vtrue{3}$ &$\vfalse{0} $ &$1$& $1$& $0$ &$3$\\
$\vtrue{3}$ & $\vtrue{1}$ &$\vfalse{0} $ &$1$& $1$& $0$ &$3$\\
$\vtrue{3}$ & $\vtrue{3}$ &$\vfalse{2} $ &$1$& $1$& $0$ &$3$\\
\end{tabular}
\end{table}

%%%%%%%%%%

\subsubsection*{Padding Bit and Headroom}
The difficulty in our approach is that for the gate types AND, NAND, OR, and NOR the corresponding function $\gateLUT{T}$ is not negacyclic. In particular, as shown in Table~\ref{tbl:id_unique}, $\gateLUT{AND}(3)=1\ne -\gateLUT{AND}(1)=0$ and $\gateLUT{OR}(0)=0 \neq -\gateLUT{OR}(2)=1$, whereas negacyclicity requires $\gateLUT{T}(i+2)=-\gateLUT{T}(i)$. Therefore, as discussed in Section~\ref{sec:lut}, a padding bit must be reserved. This bit, however, breaks the mod~$4$ arithmetic that our encoding previously provided automatically.

To address this problem, in addition to the padding bit, we reserve a \textit{headroom} in our encoding to keep the intermediate values obtained from a sequence of additions (XOR gates) away from the padding bit.
For a headroom $H>4$, each intermediate value $v\in\mathbb{Z}_H$ is encoded as $\Delta v$ with $\Delta=q/(2H)$. The factor of $2$ accounts for the reserved padding bit: the lower half of $\mathbb{Z}_q$ holds $H$ distinct values and the upper half is the padding. Additions are then performed over the integers instead of modulo $4$, and no wrap-around occurs as long as the value stays below $H$. Whenever a gate requires the non-linear function $\gateLUT{T}$, programmable bootstrapping is applied using the LUT \(x\to \gateLUT{T}(x\bmod 4)\), defined over the entire range \([0,H)\). Similarly, when the value approaches \(H\), programmable bootstrapping is applied using the LUT \(x\to \demap(x\bmod 4)\) to obtain a fresh ciphertext.

\subsubsection*{Choosing Headroom}
A larger headroom allows more XORs between bootstrappings, but it is limited by the modulus switching error $\sigma_E$ of Equation~\eqref{eq:sigma_total}.
With the padding bit the plaintext modulus is $p=2H$, so the condition $p\le N/(l\sigma_E)$ of Equation~\ref{eq:choosing_p} gives
\begin{equation}
\label{eq:headroom_max}
H\le\frac{N}{2\,l\,\sigma_E},
\end{equation}
where $l$ is chosen based on the target failure probability. Thus $H$ must be larger than the biggest value reached between two bootstrappings, and, at the same time, small enough for this bound to hold.

%%%%%%%%%%%%%%%%%
\subsection{Gate Bootstrapping for Supported Non-linear Boolean Gates}
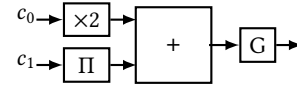
\begin{figure}[H]
\centering
\begin{tikzpicture}[ >=latex,thick, every node/.style={font=\small}]
\def\arrowTextOffset{-0.7mm}
\node[ draw,minimum width=1.cm, minimum height=1.cm] (plus) at (0,0) {$\mathsf{+}$}; % plus sing   
\node[ draw,minimum width=.6cm, minimum height=.4cm] (mult) at ([xshift=-1.15cm,yshift=.85cm]plus.south) {$\times 2$}; % x2
\draw[->] ([xshift=-.35cm]mult.west) -- (mult.west) node[midway,left,xshift=-0.05cm,yshift=.05cm] {$c_0$}; % input c0
\draw[->] (mult.east) -- ([xshift=0.35cm]mult.east) node[midway,above] {}; %x2 output
\node[ draw,minimum width=.6cm, minimum height=.4cm] (H) at ([xshift=-1.15cm,yshift=.25cm]plus.south) {$\demapHomo$}; % H
\draw[->] ([xshift=-.35cm]H.west) -- (H.west) node[midway,left,xshift=-0.05cm,yshift=.05cm] {$c_1$}; %H input c1
\draw[->] (H.east) -- ([xshift=0.35cm]H.east) node[midway,above] {}; %H output
\node[ draw,minimum width=.2cm, minimum height=.2cm] (F) at ([xshift=0.6cm]plus.east) {$\gateLUTHomo$}; % F
\draw[->] (F.east) -- ([xshift=0.35cm]F.east) node[midway,right,xshift=-.7mm] {}; %F output
\draw[->] ([xshift=-0.4cm]F.west) -- ([xshift=0cm]F.west) node[midway,above] {}; %F input    
\end{tikzpicture}
\caption{Gate bootstrapping for the supported non-free binary gates. The input
$c_0$ is multiplied by $2$, and $c_1$ is passed through the bootstrapping
$\demapHomo$ that computes $x \bmod 2$. Their sum encrypts the index
$I(m_0,m_1)$, and the bootstrapping $\gateLUTHomo$ applies the indexed truth table
of the gate $\mathsf{T}$ to it.}
\label{fig:design_ct}
\end{figure}

To evaluate a gate $\mathsf{T}$ with characteristic function $\gateLUT{T}$ on two input ciphertexts, it suffices to homomorphically evaluate $\gateLUT{T}$ as in Equation~\ref{eq:evaluation}. 
Let $\demapHomo$ and $\gateLUTHomo$ denote programmable bootstrapping operations implementing the functions $\demap(x)=x\bmod 2$ and $x\mapsto \gateLUT{T}(x\bmod 4)$, respectively. 
Given $c_0=\lwe{m_0}$ and $c_1=\lwe{m_1}$, with $\Delta=q/2H$, where $H$ is the headroom and $m_0,m_1\in\mathbb{Z}_4$ are mapped inputs, the homomorphic addition and scalar multiplication properties of TFHE yield

\begin{align*}
\gateLUTHomo\big(2c_0+\demapHomo(c_1)\big)
&=\gateLUTHomo\big(2\,\lwe{m_0}+\demapHomo(\lwe{m_1})\big)\\
&=\gateLUTHomo\big(\lwe{2m_0}+\lwe{\demap(m_1)}\big)\\
&=\gateLUTHomo\big(\lwe{(2m_0+\demap(m_1))}\big)\\
&=\gateLUTHomo\big(\lwe{I(m_0,m_1)}\big)\\
&=\lwe{\gateLUT{T}(m_0,m_1)}\\
&=\lwe{\mathsf{T}(\demap(m_0),\demap(m_1))}.
\end{align*}
The LUT of $\gateLUTHomo$ is defined on the whole range $[0,H)$ and reduces its input modulo $4$ before applying the indexed truth table. This is what makes the third line valid: the sum $2m_0+\demap(m_1)$ is computed over the integers and may exceed $3$, but $2m_0 \bmod 4$ depends only on the parity of $m_0$. It must stay below $H$, i.e., $m_0<(H-1)/2$, otherwise it spills into the padding bit. Figure~\ref{fig:design_ct} shows the resulting circuit for the supported non-free gates AND, NAND, OR and NOR (XOR and XNOR are handled by the free-XOR property, XNOR with an additional plaintext addition of $1$). 
Each such gate costs two bootstrappings, one for $\demapHomo$ and one for $\gateLUTHomo$.

The unary NOT gate is free in the same way: it is the XOR with the constant $1$, i.e., a single plaintext addition of $\Delta$ that needs no bootstrapping and adds no noise.
The constant gates FALSE and TRUE need no bootstrapping either, since their output is a trivial encryption of $0$ or $1$ that is independent of the inputs. 

%%%%%%%%%%%%%%%%%%%%%%%%%%%%%%%%%%%%%%%%%%%%% implementations %%%%%%%%%%%%%%%%%%%%%%%%%%%%%%%%%%%%%%%%%%%%%

\section{Implementations and Proof of Work}
\label{sec:implementations}
We implemented our proposed approach in Rust and made the source code publicly available in \cite{ajorian_freexor_fhe}. Our implementation uses the open-source TFHE-rs library \cite{TFHE-rs}, which provides high-level APIs for manipulating TFHE ciphertexts. To evaluate Boolean circuits, we developed a compiler that translates circuits represented in the Bristol Fashion (BF) format into equivalent circuits operating over TFHE ciphertexts. An interpreter then evaluates the resulting circuits using a valid set of ciphertexts as inputs.

\subsection{Bristol Fashion (BF)}
We use the Bristol Fashion (BF) syntax \cite{archer2019bristol} to represent Boolean circuits within our framework. BF provides a textual representation of a circuit in which all gates are listed in topological order. Each gate is specified by its type together with the indices of its input and output wires. For a circuit with $n$ input wires and $m$ output wires, the first $n$ wire indices are reserved for the input wires, while the final $m$ wire indices are reserved for the output wires.

\subsection{Parameters}
As described in Section~\ref{sec:our_encoding}, our encoding uses two bits to represent the plaintext modulus, i.e., \(p=4\). To satisfy Equation~\ref{eq:headroom_max} under the other chosen parameters, including \(N\) and \(\sigma_e\), we allocate additional two bits for carry values arising during homomorphic computations. Accordingly, we set the headroom to four bits, i.e., \(H=16\). A larger value of \(H\) allows more homomorphic operations to be performed between bootstrapping operations, thereby reducing the frequency of bootstrapping. However, increasing \(H\) may require adjusting other parameters to maintain the desired security level and may increase the dimensions of the underlying polynomials and vectors, potentially increasing computational costs. Optimizing these parameters jointly to balance security, computational efficiency, and bootstrapping frequency remains an open engineering problem for our approach.

\subsection{Lazy Bootstrapping}
As explained in Section~\ref{sec:our_encoding}, we reserve a headroom that allows us to postpone the bootstrapping of XOR, XNOR, and NOT gates, while guaranteeing that intermediate values never overwrite the padding bit, which must remain zero for the trick that enables the evaluation of non-negacyclic functions. If a value exceeds the maximum value that fits in the headroom, the padding bit is overwritten and the result is no longer valid.

One way to satisfy this requirement is to write a compiler that, given the input values, finds the optimal places to insert bootstrapping operations. However, this approach has two drawbacks: first, it may leak information about the input values, and second, it requires a new configuration for every evaluation.

Our approach instead estimates the maximum possible value (MPV) of each wire during homomorphic evaluation and applies a bootstrapping when this value approaches the headroom. To this end, we maintain a hashmap from wire indices to MPVs, which is filled and updated during the evaluation. For each gate, we look up the MPVs of its input wires and estimate the MPV of its output wire. If the estimate would overflow the headroom, we bootstrap one or two of the input wires.

The overflow condition depends on the gate type. For XOR, the evaluation is an addition of the two input ciphertexts, so an overflow occurs when the sum of the input MPVs is greater than $H-1$. For NOT, the evaluation adds 1 to the ciphertext, so an overflow occurs when the input MPV plus 1 is greater than $H\!-\!1$. For XNOR, we add the two input ciphertexts and 1, so an overflow occurs when the sum of the input MPVs plus 1 is greater than $H-1$. For the other supported symmetric gates, which apply the index function $I = 2m_0 + \pi(m_1)$, an overflow occurs when twice the MPV of the first input is greater than $H-1$. Since $\pi$ is applied via a PBS, there is no need to check the sum $2\,\mathrm{MPV}_0 + \mathrm{MPV}_1$. Table~\ref{tbl:mpvs} summarizes all cases. For XOR and XNOR, we bootstrap the input with the larger MPV first, and then the other input only if the condition still holds.

\begin{table}[t]
\centering
\setlength{\tabcolsep}{3pt}
\caption{Lazy bootstrapping conditions and estimated maximum possible value (MPV) of the output wire for each gate type, where $\mpv{i}$ is the MPV of input wire $i$ and $H$ is the headroom.}
\label{tbl:mpvs}
\begin{tabular}{c|c|c}
Gate & Bootstrapping condition & Output wire MPV \\ \hline
XOR  & $\mpv{0}+\mpv{1}>H-1$   & $\mpv{0}+\mpv{1}$ \\ \hline
XNOR & $\mpv{0}+\mpv{1}+1>H-1$ & $\mpv{0}+\mpv{1}+1$ \\ \hline
NOT  & $\mpv{0}+1>H-1$         & $\mpv{0}+1$ \\ \hline
Others & $2\,\mpv{0}>H-1$      & $2\,\mpv{0}+1$ \\
\end{tabular}
\end{table}

\subsection{Experimental Results}
% We evaluated our approach on arithmetic-operation circuits from \cite{benchbf}, cryptographic circuits from \cite{benchbf}, and benchmark circuits from \cite{benchbch}, all developed independently of our work
We evaluated our approach on toy and cryptographic circuits with varying combinations of free-XOR gates and symmetric gates requiring two PBS operations.
Each experiment was repeated 10 times on an Apple M2 MacBook Pro with 24 GB of RAM. 
Table \ref{tbl:exp_res_toys} presents selected experimental results for toy circuits, comparing the evaluation time of our approach with that of the standard approach, which performs bootstrapping after evaluating each gate. Our approach achieves an approximately \textbf{89\%} speedup on XOR-only circuits, with diminishing gains as the proportion of symmetric gates relative to FreeXOR gates increases, as expected. 
Nevertheless, our approach remains effective for circuits with a higher proportion of symmetric gates. In particular, the cryptographic circuits listed in Table~\ref{tbl:exp_res_crypto} demonstrate speedups of \textbf{30\%} for AES-128 and AES-192, \textbf{23\%} for AES-256, and \textbf{34\%} for both SHA-256 and SHA-512.

\begin{table}[H]
\centering
\setlength{\tabcolsep}{3pt}
\caption{Evaluation times of our approach compared with the standard approach on toy circuits with varying proportions of Free-XOR and symmetric gates. All the elapsed times are reported in seconds.}
\label{tbl:exp_res_toys}
\begin{tabular}{c|c|c|c}
Circuit & Gates & Standard & FreeXOR \\ \hline
XORTree & 63 XOR & 1.180 & 0.130 \\ \hline
toy1 & 15 OR, 150 XOR, 20 AND, 15 INV & 3.719 & 1.688 \\ \hline
toy2 & 30 OR, 30 INV, 100 XOR, 40 AND & 3.708 & 2.966 \\ \hline
toy3 & 80 XOR, 20 INV, 50 AND, 50 OR & 3.708 & 4.515 \\ \hline
\end{tabular}
\end{table}

\begin{table}[H]
\centering
\setlength{\tabcolsep}{3pt}
\caption{Evaluation times of our approach compared with the standard approach on cryptographic circuits, including AES and SHA variants. All the elapsed times are reported in seconds.}
\label{tbl:exp_res_crypto}
\begin{tabular}{c|c|c|c}
Circuit & Gates & Standard & FreeXOR \\ \hline
AES128& 28176 XOR, 6400 AND, 2087 INV & 694.730 & 480.198 \\ \hline
AES192&7168 AND, 2317 INV, 32080 XOR &789.927 &540.505 \\ \hline
AES256& 8832 AND, 39008 XOR, 2826 INV & 962.746 & 736.145\\ \hline 
SHA256 & 22573 AND, 110644 XOR, 1856 INV & 2525.679 & 1667.625 \\ \hline
SHA512 & 4946 INV, 57947 AND, 286724 XOR & 6448.118 & 4219.999 \\ \hline
\end{tabular}
\end{table}

 \section{Conclusion and Discussion}
\label{sec:conc}

We proposed an encoding that enables XOR evaluation without immediate programmable bootstrapping (PBS), using homomorphic addition instead. To support other symmetric gates, such as AND, NAND, OR, and NOR, we introduced an indexing function that homomorphically identifies the corresponding row of a gate's truth table. Evaluating a symmetric gate requires two PBS operations: one to compute the index and another to evaluate the gate's functionality based on that index. In contrast, XOR gates can be evaluated without immediate bootstrapping. Our approach reserves headroom for intermediate results produced by operations performed without bootstrapping, allowing bootstrapping to be deferred until this headroom is about to overflow. Consequently, our approach is most efficient for circuits in which XOR gates significantly outnumber symmetric gates.

%\begin{acks}

%\end{acks}

\bibliographystyle{ACM-Reference-Format}
\bibliography{main}

@inproceedings{cheon2017homomorphic,
  title={Homomorphic encryption for arithmetic of approximate numbers},
  author={Cheon, Jung Hee and Kim, Andrey and Kim, Miran and Song, Yongsoo},
  booktitle={International conference on the theory and application of cryptology and information security},
  pages={409--437},
  year={2017},
  organization={Springer}
}

@article{brakerski2014leveled,
  title={(Leveled) fully homomorphic encryption without bootstrapping},
  author={Brakerski, Zvika and Gentry, Craig and Vaikuntanathan, Vinod},
  journal={ACM Transactions on Computation Theory (TOCT)},
  volume={6},
  number={3},
  pages={1--36},
  year={2014},
  publisher={ACM New York, NY, USA}
}

@inproceedings{ducas2015fhew,
  title={FHEW: bootstrapping homomorphic encryption in less than a second},
  author={Ducas, L{\'e}o and Micciancio, Daniele},
  booktitle={Annual international conference on the theory and applications of cryptographic techniques},
  pages={617--640},
  year={2015},
  organization={Springer}
}

@article{chen2024free,
  title={Free-xor gate bootstrapping},
  author={Chen, Chunling and Lu, Xianhui and Wang, Ruida and Li, Zhihao and Shen, Xuan and Wei, Benqiang},
  journal={Cryptology ePrint Archive},
  year={2024}
}

@phdthesis{gentry2009fullyphdthesis,
  title={A fully homomorphic encryption scheme},
  author={Gentry, Craig},
  year={2009},
  school={Stanford University}
}

@misc{chillotti2022tfhedeepdive,
  author       = {Ilaria Chillotti},
  title        = {{TFHE Deep Dive - Part IV - Programmable Bootstrapping}},
  year         = {2022},
  month        = jun,
  howpublished = {\url{https://www.zama.org/post/tfhe-deep-dive-part-4}},
  note         = {Zama Blog}
}

@article{joye2021guide,
  title={Guide to fully homomorphic encryption over the [discretized] torus},
  author={Joye, Marc},
  journal={Cryptology ePrint Archive},
  year={2021}
}

@article{brakerski2014efficient,
  title={Efficient fully homomorphic encryption from (standard) LWE},
  author={Brakerski, Zvika and Vaikuntanathan, Vinod},
  journal={SIAM Journal on computing},
  volume={43},
  number={2},
  pages={831--871},
  year={2014},
  publisher={SIAM}
}

@article{chillotti2020tfhe,
  title={TFHE: Fast Fully Homomorphic Encryption Over the Torus: I. Chillotti et al.},
  author={Chillotti, Ilaria and Gama, Nicolas and Georgieva, Mariya and Izabach{\`e}ne, Malika},
  journal={Journal of Cryptology},
  volume={33},
  number={1},
  pages={34--91},
  year={2020},
  publisher={Springer}
}

@inproceedings{rothblum2011homomorphic,
  title={Homomorphic encryption: From private-key to public-key},
  author={Rothblum, Ron},
  booktitle={Theory of cryptography conference},
  pages={219--234},
  year={2011},
  organization={Springer}
}

@article{rivest1978data,
  title={On data banks and privacy homomorphisms},
  author={Rivest, Ronald L and Adleman, Len and Dertouzos, Michael L and others},
  journal={Foundations of secure computation},
  volume={4},
  number={11},
  pages={169--180},
  year={1978}
}

@article{rivest1978method,
  title={A method for obtaining digital signatures and public-key cryptosystems},
  author={Rivest, Ronald L and Shamir, Adi and Adleman, Len},
  journal={Communications of the ACM},
  volume={21},
  number={2},
  pages={120--126},
  year={1978},
  publisher={ACM New York, NY, USA}
}

@inproceedings{boneh2005evaluating,
  title={Evaluating 2-DNF formulas on ciphertexts},
  author={Boneh, Dan and Goh, Eu-Jin and Nissim, Kobbi},
  booktitle={Theory of cryptography conference},
  pages={325--341},
  year={2005},
  organization={Springer}
}

@article{elgamal1985public,
  title={A public key cryptosystem and a signature scheme based on discrete logarithms},
  author={ElGamal, Taher},
  journal={IEEE transactions on information theory},
  volume={31},
  number={4},
  pages={469--472},
  year={1985},
  publisher={IEEE}
}

@article{regev2009lattices,
  title={On lattices, learning with errors, random linear codes, and cryptography},
  author={Regev, Oded},
  journal={Journal of the ACM (JACM)},
  volume={56},
  number={6},
  pages={1--40},
  year={2009},
  publisher={ACM New York, NY, USA}
}

@inproceedings{chillotti2016faster,
  author    = {Chillotti, Ilaria and Gama, Nicolas and Georgieva, Mariya and Izabach{\`e}ne, Malika},
  title     = {Faster Fully Homomorphic Encryption: Bootstrapping in Less Than 0.1 Seconds},
  booktitle = {Advances in Cryptology -- ASIACRYPT 2016},
  year      = {2016},
  publisher = {Springer},
  pages     = {3--33},
  doi       = {10.1007/978-3-662-53887-6_1}
}

@article{acar2018survey,
  title={A survey on homomorphic encryption schemes: Theory and implementation},
  author={Acar, Abbas and Aksu, Hidayet and Uluagac, A Selcuk and Conti, Mauro},
  journal={ACM Computing Surveys (Csur)},
  volume={51},
  number={4},
  pages={1--35},
  year={2018},
  publisher={ACM New York, NY, USA}
}

@article{marcolla2022survey,
  title={Survey on fully homomorphic encryption, theory, and applications},
  author={Marcolla, Chiara and Sucasas, Victor and Manzano, Marc and Bassoli, Riccardo and Fitzek, Frank HP and Aaraj, Najwa},
  journal={Proceedings of the IEEE},
  volume={110},
  number={10},
  pages={1572--1609},
  year={2022},
  publisher={IEEE}
}

@software{ajorian_freexor_fhe,
  author  = {Ali Ajorian},
  title   = {{FreeXOR-FHE}},
  url     = {https://github.com/Ajorian/freexor-fhe},
  year    = {2026},
  note    = {Source code repository}
}

@misc{archer2019bristol,
  author       = {Archer, David and Arribas Abril, Victor and Lu, Steve and Maene, Pieter and Mertens, Nele and Sijacic, Danilo and Smart, Nigel},
  title        = {{Bristol Fashion} MPC Circuits},
  year         = {2019},
  howpublished = {\url{https://nigelsmart.github.io/MPC-Circuits/}},
  note         = {Accessed: 2026-09-10}
}

@misc{TFHE-rs,
  title  = {{TFHE-rs: A Pure Rust Implementation of the TFHE Scheme for Boolean and Integer Arithmetics Over Encrypted Data}},
  author = {Zama},
  year   = {2022},
  note   = {\url{https://github.com/zama-ai/tfhe-rs}},
}

\appendix

%massa et mattis lacinia.

\end{document}